\documentclass[twocolumn]{aastex61}

\usepackage{graphicx}
\usepackage{color}
\usepackage{epstopdf}

\begin{document}

\newcommand{\Msun}{\ensuremath{{\rm M}_\odot}}                    
\newcommand{\Rsun}{\ensuremath{{\rm R}_\odot}}                    
\newcommand{\Lsun}{\ensuremath{{\rm L}_\odot}}                    
\newcommand{\psun}{\ensuremath{\rho_\odot}}                       
\newcommand{\Mjup}{\ensuremath{{\rm M}_{\rm Jup}}}                
\newcommand{\Rjup}{\ensuremath{{\rm R}_{\rm Jup}}}                
\newcommand{\pjup}{\ensuremath{\rho_{\rm Jup}}}                   
\newcommand{\Mearth}{\ensuremath{\,{\rm M}_\oplus}}               
\newcommand{\Rearth}{\ensuremath{\,{\rm R}_\oplus}}               
\newcommand{\Teff}{\ensuremath{{\rm T}_{\rm eff}}}                
\newcommand{\FeH}{\ensuremath{\left[{\rm Fe/H}\right]}}           
\newcommand{\as}{\ensuremath{^{\prime\prime}}}                    
\newcommand{\am}{\ensuremath{^\prime}}                            
\newcommand{\degrees}{\ensuremath{^\circ}}                        
\newcommand{\Rstar}{\ensuremath{R_\star}}                         
\newcommand{\Mstar}{\ensuremath{M_\star}}                         
\newcommand{\Lstar}{\ensuremath{{\rm L}_\star}}                   
\newcommand{\pstar}{\ensuremath{\rho_\star}}                      
\newcommand{\EBV}{\ensuremath{\textrm{E}\left(\textrm{B}-\textrm{V}\right)}}
\newcommand{\Ks}{\ensuremath{\textrm{K}_{\,\textrm{s}}}}
\newcommand{\Teq}{\ensuremath{{\rm T}_{\rm eq}^{\,\prime}}}       
\newcommand{\safronov}{\ensuremath{\Theta}}                       
\newcommand{\mss}{\,m\,s$^{-2}$}                                  
\newcommand{\mcc}[1]{\multicolumn{3}{c}{#1}}
\newcommand{\ermcc}[5]{\mcc{\ensuremath{{#1\,^{+#2}_{-#3}}\,^{+#4}_{-#5}}}}
\newcommand{\ercc}[3]{\mcc{\ensuremath{#1^{+#2}_{-#3}}}}

\submitjournal{ApJL}

\title{WASP-20 is a close visual binary with a transiting hot Jupiter}

\author{Daniel F. Evans}
\affil{	Astrophysics Group, Keele University, Staffordshire, ST5 5BG, UK }

\author{John Southworth}
\affil{	Astrophysics Group, Keele University, Staffordshire, ST5 5BG, UK }

\author{Barry Smalley}
\affil{	Astrophysics Group, Keele University, Staffordshire, ST5 5BG, UK }

\correspondingauthor{Daniel F. Evans}
\email{d.f.evans@keele.ac.uk}

\begin{abstract}
We announce the discovery that WASP-20 is a binary stellar system, consisting of two components separated by $0.2578\pm0.0007\as$\ on the sky, with a flux ratio of $0.4639\pm 0.0015$ in the $K$-band. It has previously been assumed that the system consists of a single F9\,V star, with photometric and radial velocity signals consistent with a low-density transiting giant planet. With a projected separation of approximately $60$\,au between the two components, the detected planetary signals almost certainly originate from the brighter of the two stars. We reanalyse previous observations allowing for two scenarios, `planet transits A' and `planet transits B', finding that both cases remain consistent with a transiting gas giant. However, we rule out the `planet transits B' scenario because the observed transit duration requires star B to be significantly evolved, and therefore have an age much greater than star A. We outline further observations which can be used to confirm this finding. Our preferred `planet transits A' scenario results in the measured mass and radius of the planet increasing by 4$\sigma$ and 1$\sigma$, respectively.
\end{abstract}


\section{Introduction} \label{sec:intro}

Contaminating light can be a significant source of systematic errors in the study of transiting exoplanets, with both photometric and spectrosopic measurements being affected. The observed depth of a transit is reduced and radial velocity signals are diluted, causing the orbiting object to appear smaller and less massive (e.g.\ \citealt{2009A&A...498..567D}, \citealt{2011ApJS..197....3B}). As the contamination typically has a wavelength dependence, the results of transmission photometry and spectroscopy are also affected, as these techniques are based on detecting the wavelength dependence of the transit depth. Where the fraction of contaminating light is high, sources such as eclipsing binaries can mimic planetary transits \citep{2003ApJ...593L.125B}. As a result, the false positive rate of photometric surveys is very high; the WASP survey rejects approximately 90\% of candidates \citep{2011EPJWC..1101004H}, whilst a third of Kepler hot Jupiter candidates are likely to be false positives \citep{2012A&A...545A..76S}.

Our ongoing High-resolution Imaging of Transiting Extrasolar Planetary systems (HITEP) survey \citep{2016A&A...589A..58E} aims to detect and characterise companion stars to known hot Jupiter host stars. With knowledge of these companion stars, it is then possible to correct for the contamination present in observations of the planet. We also aim to better characterise the multiplicity of such systems, as binary companions are likely to have an important influence on hot Jupiter formation and evolution (e.g.\ \citealt{1996Sci...274..954R, 2014ApJ...783....4W}).

In this paper, we present the discovery that the WASP-20 system is a close visual binary. This system was previously characterised by \citet[][hereafter A15]{2015A&A...575A..61A} as a single late-F main sequence star hosting a transiting `hot Saturn' in a 4.9-day orbit. Line bisector analysis did not reveal any potential complications, and no long-term radial velocity (RV) signals were found across five years of data. It was noted that WASP-20\,b has a very low surface gravity and density, with $\rho_p=0.099\pjup$, joining other extremely low density objects such as WASP-17 \citep{2010ApJ...709..159A,2012MNRAS.426.1338S} and TrES-4 \citep{2015A&A...575L..15S}. Combined with a relatively bright host star ($V=10.7$), WASP-20\,b therefore appeared to be a tempting target for atmospheric characterisation.

We previously observed WASP-20 in 2014 using the Two Colour Instrument, a lucky imaging instrument on the 1.5\,m Danish Telescope at La Silla, but did not resolve any companion stars. In the best 1\% of lucky imaging exposures, the stellar point spread function has a full width at half maximum of $0.46\as$ \citep{2016A&A...589A..58E}. We present new photometric and spectroscopic observations taken with VLT/SPHERE, clearly resolving WASP-20 as two separate sources separated by $0.26\as$, which we designate WASP-20\,A and WASP-20\,B. We re-analyse published photometry and spectroscopy, accounting for the binarity of the host star, and demonstrate the implications for the planet.

\section{Observations and data reduction}

We obtained photometric and low-resolution spectroscopic observations of WASP-20 on the night of 2016/11/05 with the SPHERE instrument \citep{2008SPIE.7014E..18B} mounted on the VLT. Data were obtained simultaneously using the InfraRed Dual-beam Imager and Spectrograph (IRDIS, \citealt{2008SPIE.7014E..3LD}) and the infrared Integral Field Spectrograph (IFS, \citealt{2008SPIE.7014E..3EC}). We also re-analyse the HARPS \citep{2003Msngr.114...20M} and CORALIE \citep{2000A&A...354...99Q} spectra, and the EulerCam \citep{2012A&A...544A..72L} $r$-band light curves of WASP-20 published in A15.

A dichroic was used to split light between IRDIS and the IFS, with IRDIS being operated in Classical Imaging mode and receiving light in the $K$-band, whilst the IFS was operated in $YJH$ mode to give spectra with $R\simeq30$ in the range $0.95$--$1.65\mu$m. The $\mathrm{N\_ALC\_YJH\_S}$ coronograph was used to block light from the primary star in order to search for faint companions.

\begin{figure*}[t]
	\plotone{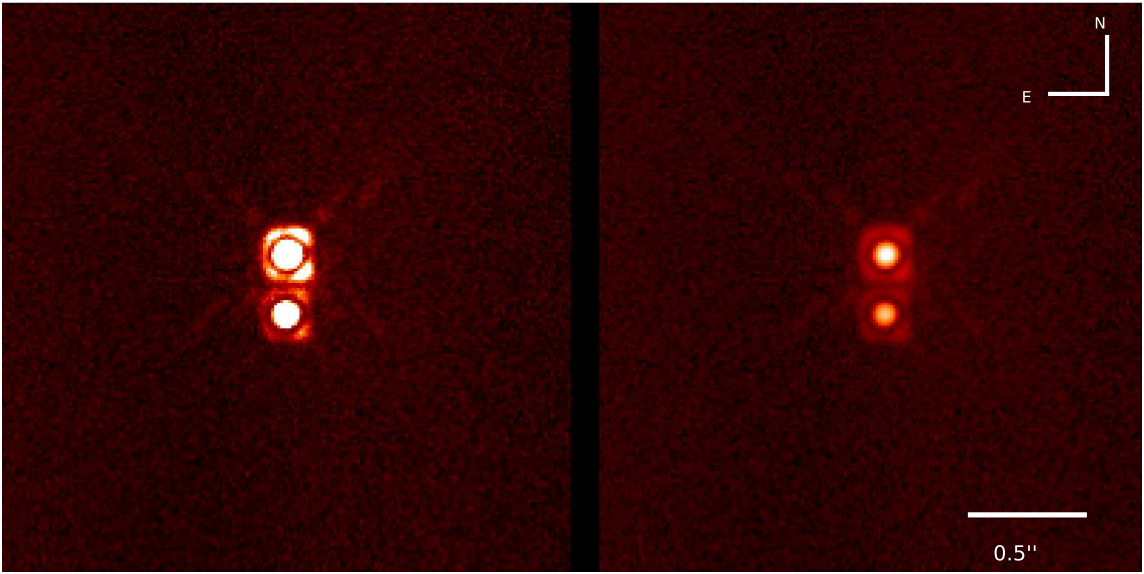}
	\caption{A $2.5\as\times2.5\as$ section of a non-coronographic IRDIS K-band image, shown with both a linear colour scale (left) and a logarithmic colour scale (right). The images clearly show the binary nature of WASP-20. We denote the upper, brighter component as WASP-20\,A, and the lower, fainter component as WASP-20\,B. \label{fig:IRDISImage}}
\end{figure*}

Observations were carried out using a flux-centre-object sequence. A flux calibration frame was obtained without the coronograph centred on the target star and taking a short exposure of the unobscured target. The target was then centred behind the coronograph and a target centring frame obtained, following the principles outlined in \citet{2006ApJ...647..612M} and \citet{2006ApJ...647..620S}. The target was then observed for a total exposure time of 256\,s to search for faint companions, using 4\,s exposures for IRDIS and 8\,s exposures for the IFS. All calibrations used to reduce the observations were taken as part of the standard SPHERE calibration sequence.

The data were reduced using version 0.18.0 of the SPHERE reduction pipeline\footnote{\url{http://www.eso.org/sci/software/pipelines/sphere/sphere-pipe-recipes.html}}. For IRDIS images, the instrumental background was subtracted, the flat field divided out, and correction made for detector distortion. For IFS data, a dark frame was subtracted, the wavelength-dependent flat field divided out, and correction made for detector distortion. Positions of the individual spectra on the detector were calibrated by illuminating the instrument's lenset array with a broadband lamp, and then wavelength-calibrated by illuminating the array with four lasers. The individual spectra were extracted and resampled into a cube of 39 monochromatic images on a $7.4$\,mas$^2$ grid.

We noticed that the telluric absorption features at $1.1\mu$m and $1.4\mu$m in the IFS spectra appeared at slightly redder wavelengths than expected. \citet{2015MNRAS.454..129V} previously noted similar offsets in the wavelength calibration, finding that the SPHERE reduction pipeline's wavelength solution gave offsets of up to $20$\,nm for the calibration laser wavelengths. We therefore re-derived the wavelength calibration using a similar method to \citet{2015MNRAS.454..129V}, by reducing the laser-illuminated wavelength calibration frame the pipeline as if it were a standard imaging frame. We then fitted the four laser line positions with Gaussian profiles, and created a second-order polynomial wavelength solution via least squares fitting.

We performed astrometric calibration for IRDIS using observations of the globular cluster 47 Tucanae taken on 2016/09/20. The data were reduced as standard IRDIS images, and the positions of stars were determined using \texttt{DAOPHOT} \citep{1987PASP...99..191S}. The measured positions were then matched to Hubble Space Telescope (HST) measurements of 47 Tuc \citep{2006ApJS..166..249M}, with our measured stars matched with the closest reference star and the offset between the pair calculated. The sum of the offsets was minimised using a four-parameter fit to detector scale, detector rotation, and x/y offsets. Our measured stellar positions were corrected for first-order effects in detector distortion by multiplying the measured vertical pixel positions by a factor of $1.0060\pm0.0002$ \citep{2016arXiv160906681M}.

We derived a detector scale of $12.301\pm0.032$\,mas\,px$^{-1}$ and a rotation of $1.81\pm0.07$\degrees\ westwards of North. Our detector scale is $1.1\sigma$ higher than that found by \citet{2016arXiv160906681M}, possibly due to changes in stellar positions since the reference data epoch (2002.26), which cannot be corrected for in the \citet{2006ApJS..166..249M} dataset as proper motion information is not available for the targeted field. This problem will be solved in the near future when the proper motion data used by \citet{2016arXiv160906681M} becomes publicly available \citep{2014ApJ...797..115B}.

\begin{figure*}[t]
	\plotone{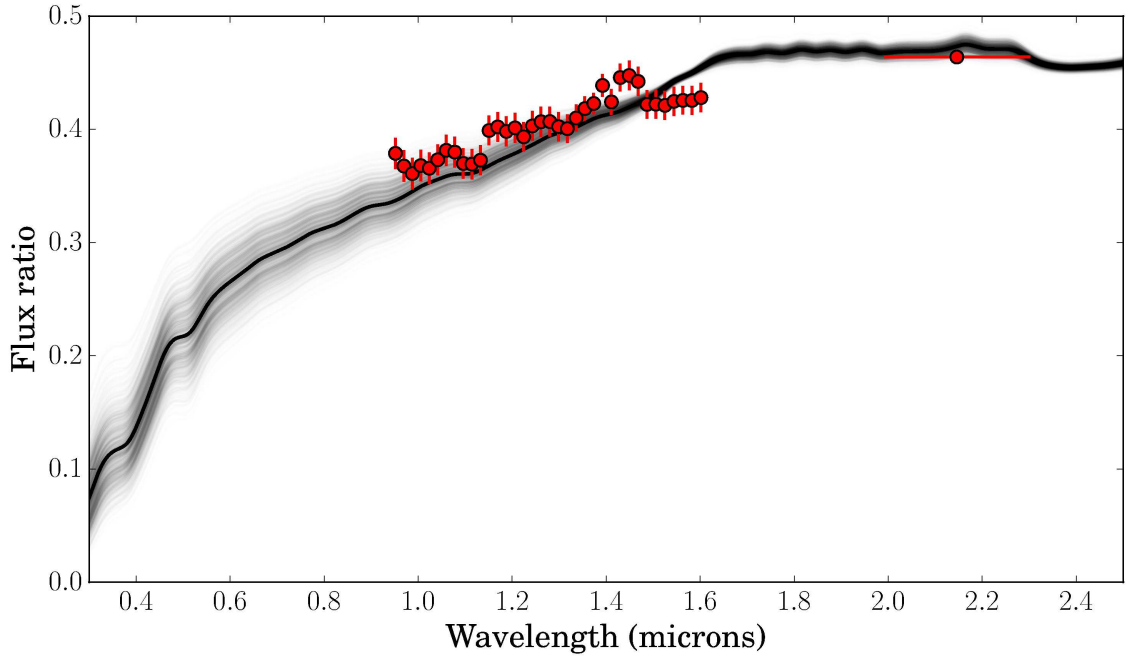}
	\caption{The measured flux ratios of the two stars from IRDIS and IFS data. The black line indicates the best fit flux ratio, based on {\sc atlas9} model atmospheres at $T_1=6000\pm100$\,K and $T_2=5060\pm250$\,K. The light grey lines show individual solutions, with the input parameters randomly perturbed within their uncertainties. \label{fig:fluxratiosfit}}
\end{figure*}

\section{Results}

The IRDIS images clearly resolve WASP-20 as two stars, which we denote A (brighter) and B (fainter), with the IFS providing a resolved spectrum for each star. We measure a $K$-band flux ratio of $0.4639\pm 0.0015$ ($\Delta\mathrm{K}=0.834\pm 0.004$\,mag) from the IRDIS data. From the IFS data we derive a $H$-band flux ratio of $0.425\pm0.012$ ($\Delta\mathrm{H}=0.93\pm0.03$ mag.) and a $J$-band flux ratio of $0.374\pm0.014$ ($\Delta\mathrm{J}=1.07\pm0.04$\,mag), although we note that the IFS data cover only the lower half of the $H$-band. Based on the ephemeris and period given in A15, these flux ratios were measured out of transit.

The decreasing flux ratio towards optical wavelengths indicate that WASP-20\,B is redder than WASP-20\,A, as expected for two bound main sequence stars. The HARPS data show no indication of binarity, so WASP-20\,B likely has a similar spectral type and RV to WASP-20\,A. From the IRDIS data, we measure a separation of $257.8\pm0.7$\,mas and a position angle of $176.93\pm0.07$\degrees, eastwards of North.

Our single-epoch observations do not conclusively prove that the two components are bound -- it is possible that WASP-20\,B is a background red giant or a foreground dwarf. We calculated the probability of such a chance alignment, as well as the probability of observing a binary companion at the separation and magnitude ratio measured, using the methods outlined in \citet{2016A&A...589A..58E} and \citet{2016MNRAS.463...37S}. Background contamination was modelled using v1.6 of the TRILEGAL galaxy model \citep{2005A&A...436..895G} with galactic coordinates $l=55.9, b=-82.4$, whilst the binary population model was based on the results of \citet{2010ApJS..190....1R}.

We find that the probability of a chance alignment at $257.8$\,mas or less with the correct $K$-band flux ratio is $2.0\times10^{-9}$, compared to a probability of $2.6\times10^{-4}$ for observing a binary companion at the given separation and flux ratio. With the binary model being five orders of magnitude more likely, we conclude that the two stars are bound. With a proper motion of approximately $21$\,mas\,yr$^{-1}$ \citep{2004AAS...205.4815Z, 2003AJ....125..984M, 2010AJ....139.2440R}, future observations of WASP-20 with SPHERE or similar instruments should easily be able to detect common proper motion between the two components.

To determine the temperature of WASP-20\,B, we modelled the IRDIS and IFS flux ratios using {\sc atlas9} model atmospheres \citep{2004astro.ph..5087C}. Both stars were assumed to be on the main sequence with $\log g=4.5$. The model atmospheres were scaled to the stellar surface areas using the empirical temperature-radius relationship presented in \citet{2016A&A...589A..58E}. To constrain our model, we assumed that the effective temperature given in A15, derived from the combined HARPS spectra of both objects, was minimally affected by WASP-20\,B. With our adopted temperature for WASP-20\,A of $T_A=6000\pm100$\,K, we derive a temperature of $T_B=5060\pm250$\,K for WASP-20\,B, with the fitted flux ratio shown in Fig.\,\ref{fig:fluxratiosfit}.

\begin{figure*}[t]
	\plotone{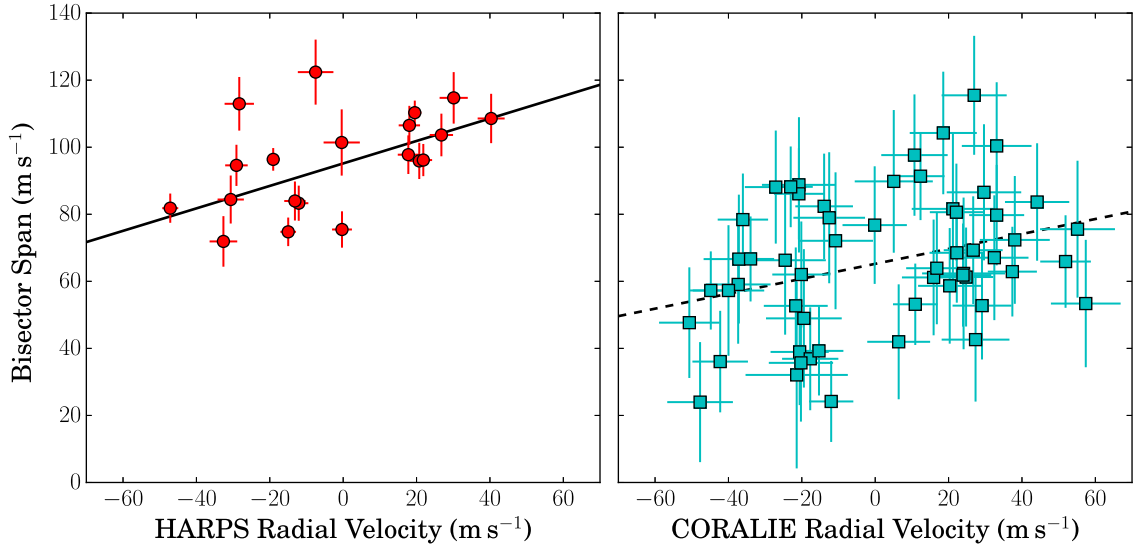}
	\caption{The measured radial velocities and bisector spans for the HARPS (left panel, red circles) and CORALIE (right panel, blue squares) observations of WASP-20. Both datasets are positively correlated, the CORALIE data having a Pearson correlation coefficient $r=0.24$, and the HARPS data having $r=0.47$. Overplotted are linear fits to the two datasets, with HARPS shown by the solid line and CORALIE by the dashed line. \label{fig:bisectors}}
\end{figure*}

\section{Planet properties}

A15 derived a distance of $210\pm20$\,pc to WASP-20, under the assumption that it was a single star. However, the apparent magnitude of WASP-20\,A will have been overestimated due to the contaminating light from WASP-20\,B. Using a flux ratio of $0.24$ in the visible and scaling the measured flux of the star, we derive an updated distance of $235$\,pc. Combined with an angular separation of $0.26\as$, the projected separation of the two components is therefore $61$\,au, ruling out a P-type circumbinary orbit for the planet. It is therefore likely that the periodic photometric and spectroscopic signals originate from a planet in an S-type orbit around one the two stars; our SPHERE data do not reveal which star this is. We therefore considered scenarios in which the planet can be hosted by either of the two stars, and calculated sets of properties for `planet transits A' and `planet transits B'. 

Of the lightcurves presented in A15, we considered only the Gunn-$r$ data, due to its low scatter and complete coverage of multiple transits. To correct for photometric dilution, we convolved our fitted flux ratios with an $r$-band filter curve\footnote{\url{http://www.not.iac.es/instruments/filters/filters.php}}, from which we derived a flux ratio of $0.269\pm0.017$ in $r$. 

\subsection{Radial velocity dilution}

In A15, radial velocities were measured by computing the cross-correlation functions of the observed HARPS and CORALIE data with a synthetic G2 spectral template. The dilution of radial velocities is not necessarily linear with flux ratio, with the profiles of individual spectral lines also varying between the two stars (e.g.\ \citealt{2015MNRAS.451.2337S}). We therefore modelled the effects of spectroscopic dilution by combining pairs of synthetic spectra representing the A and B components. The spectra were generated using \texttt{iSpec} \citep{2014A&A...569A.111B} with $R=120000$ covering $380$--$690$\,nm, representing the spectra produced by the HARPS spectrograph. The properties used for the stars were temperatures of $T_A=6000\pm100$\,K and $T_B=5060\pm250$\,K, a metallicity of $\FeH=-0.01$ (A15), a surface gravity of $\log g=4.5$, and a projected surface rotation of $v\sin i=2$ for both stars. The reference G2 template was generated with the same parameters, with the exception of $T=5770$\,K and $\FeH=0.0$.

For each iteration of the model, the input RV offset was applied to either star A or star B, and the spectra were then combined, weighted by their flux ratio at each wavelength. The combined spectrum was then cross-correlated with the template spectrum. To determine the RV dilution, the RV shift was iterated until the measured RV matched the observed value of $32.8$\,m\,s$^{-1}$ given in A15. We find that the measured RVs, $\mathrm{RV}_\mathrm{meas}$, are diluted compared to the intrinsic RVs, $\mathrm{RV}_\mathrm{0}$, by a factor of $\mathrm{RV}_\mathrm{meas} = 0.73\pm0.03 \mathrm{RV}_\mathrm{0}$ in the case that star A is the source, and by $\mathrm{RV}_\mathrm{meas} = 0.18\pm0.02 \mathrm{RV}_\mathrm{0}$ for star B, giving corrected velocity amplitudes of $44.0\pm2.3$\,m\,s$^{-1}$ and $177\pm21$\,m\,s$^{-1}$, respectively.

We do not consider in detail the implications for the Rossiter-McLaughlin observations in A15, as the dilution is sensitive to the spectroscopic properties of the individual stars, in particular their rotational velocities \citep{2015MNRAS.451.2337S}, which we do not consider here. We note that A15 state that no significant correlation exists between the measured RVs and bisector spans, but this claim is not quantified in any way, nor is it typical to do so in the literature. We calculated the Pearson correlation coefficient $r$ for both the CORALIE and HARPS datasets, and calculated confidence intervals by permuting the measured values within their error bars. Both datasets are positively correlated with high significance; the CORALIE dataset has $r=0.24$, significant at $p=0.003$, whilst the HARPS dataset has $r=0.47$, significant at $p=7.6\times10^{-8}$. If these correlations are due to WASP-20's binarity, a more detailed analysis of the bisector spans of other planetary systems may therefore detect other similar systems. We show linear fits to the data in Fig.~\ref{fig:bisectors}.

\subsection{JKTEBOP modelling}

The corrected transit photometry, RVs, and estimated stellar effective temperatures were re-analysed using the \textsc{jktebop} code \citep{2013A&A...557A.119S} using the methodology outlined in \citet{2012MNRAS.426.1291S}, assuming an eccentricity of 0 (A15). In addition to the `planet transits A' and `planet transits B' scenarios, we also modelled the system ignoring its binary, to allow for comparison with the results of A15. All results, including those from A15, are shown in Table~\ref{tab:jktebop}. Comparing the planet's parameters from the `ignoring binarity' and the `planet transits A' scenarios, the main effect is a $30 \pm 8$\% ($4\sigma$) increase in the planet's mass when RV dilution is corrected for, but with the density of the planet still being notably low. The changes are much larger when comparing to the `planet transits B' scenario, which gives a planetary mass four times larger than the `ignoring binarity' model, and a $41 \pm 16$\% ($3.5\sigma$) change in planetary radius. The change in radius for this scenario is smaller than one might expect, due to the trade-off between reducing stellar radius and increasing transit depth required to give the same depth when diluted. The planet's properties remain within the planetary regime for both scenarios.

One notable feature of the `planet transits B' results is the predicted stellar age, which is very large due to the need for the star to be significantly evolved in order to reproduce the transit duration. The star's age is much larger than that of the hotter and more luminous star A, which is problematic if they are co-evolutionary, and also exceeds the age of the Universe. Due to this problem, we conclude that the `planet transits A' scenario is strongly preferred. The planet should therefore be referred to as WASP-20\,Ab.

\section{Summary and Conclusions}

We have resolved the planet host star WASP-20 into a binary system separated by 0.26\as. The probability that they are gravitationally bound greatly exceeds that of a chance alignment. The existence of the second star caused previous measurements of the system properties to be erroneous, so we have redetermined the properties for both the `planet transits A' and `planet transits B' scenarios using published photometry and spectroscopy. The discrepant age implied by the latter option allows us to conclude that the `planet transits A' scenario is correct. We confirm the planetary nature of WASP-20\,Ab and find that its measured mass and radius are increased by $30 \pm 8$\% and $7 \pm 13$\% versus an analysis which does not account for the binarity of the host star. Our measurements remain consistent with it having a low surface gravity and density, so it is still a good candidate for transmission spectroscopic studies if the flux contribution from the second star can be corrected for.

Further observations of the system would be useful to confirm our results. This could be achieved by on-off observations, in which the flux ratio of the stars is measured both during and outside transit, revealing which of the two stars has been dimmed by the transit event. The two stars must be spatially resolved for this work, requiring the use of adaptive optics, lucky imaging, or a space telescope. Alternatively, unresolved multicolour photometry of the system would be able to detect changes of the transit depth caused by wavelength-dependent contamination. The transits will appear deeper towards the blue if the planet orbits star A, and shallower if it transits star B. Predictions for transit depths in either scenario are given in Table~\ref{tab:fluxRatios}, based on our best-fit flux ratios. We also encourage further high resolution spectroscopic observations of the two components, in order to better characterise both stars and the effect of dilution. We note in particular that our RV corrections rely on several assumptions, and that more detailed analysis of the spectroscopic dilution is required.

It is interesting to consider the implications for hot Jupiter formation and migration, considering WASP-20AB's projected separation of only $61$au and a relatively high stellar mass ratio, compared to other systems that have only very distant or low mass companions. (e.g. \citealt{2015ApJ...800..138N, 2016ApJ...827....8N}). It is likely that hot Jupiters did not form in situ, instead migrating in from orbits at several au. It has been proposed that wide stellar binaries could be a migration pathway, with the Kozai-Lidov (KL) mechanism forcing the planet into an eccentric orbit that is then shrunk by tidal friction. Following \citet{2007ApJ...669.1298F} and assuming a circular binary orbit with $a=61$au, the KL mechanism would be operative for initial planetary orbits with $a_{\rm init}>2.64$au for `planet transits A', and $a_{\rm init}>2.10$ for `planet transits B'.
	
However, if the KL mechanism operates on the planet, it may also operate on dust particles and planetesimals in the protoplanetary disc, disrupting planet formation -- this could imply that the stellar and planetary orbits are co-planar, or that the stellar orbit is much wider than the projected separation. Theoretical studies currently show little agreement regarding the effect of binary companions on planet formation; for example, \citet{2011A&A...528A..40F} predict that discs will be significantly disrupted by inclined binary companions at less than 100au, in contrast to simulations by \citet{2015ApJ...798...70R} indicating that massive planet formation can occur even in 20au binaries.

\acknowledgements{D.F.E. is funded by the UK's Science and Technology Facilities Council. J.S. acknowledges support from the Leverhulme Trust in the form of a Philip Leverhulme prize. Based on observations made with ESO Telescopes at the La Silla Paranal Observatory under programme ID 098.C-0589. This research has made use of the SIMBAD database and the VizieR catalogue access tool, operated at CDS, Strasbourg, France. This research has made use of NASA's Astrophysics Data System. }
\facility{VLT:Melipal (SPHERE)}

\begin{longrotatetable}
	\begin{table}
		\centering
		\caption{Derived physical properties for the three scenarios considered -- ignoring binarity, the planet orbiting star A, and the planet orbiting star B. The original results of A15 are provided for comparison. Where two sets of errorbars are given, the first is the statistical uncertainty and the second is the systematic uncertainty.} \label{tab:jktebop}
		\begin{tabular}{l l r@{\,$\pm$\,}c@{\,$\pm$\,}l r@{\,$\pm$\,}c@{\,$\pm$\,}l r@{\,$\pm$\,}c@{\,$\pm$\,}l r@{\,$\pm$\,}l }
			\tablewidth{0pt}
			\hline
			\hline
			Parameter  & Symbol & \multicolumn{3}{c}{Ignoring binarity} & \multicolumn{3}{c}{Planet transits A} & \multicolumn{3}{c}{Planet transits B} & \multicolumn{2}{c}{A15}  \\
			           &        & \multicolumn{3}{c}{}                  & \multicolumn{3}{c}{Adopted solution}  \\
			\hline
                  Linear limb darkening coefficient    & $u_{\rm A}$         & \mcc{$0.57 \pm 0.15$}            & \mcc{$0.59 \pm 0.15$}            & \mcc{$0.55 \pm 0.14$}                    \\
                  Quadratic limb darkening coefficient & $u_{\rm A}$         & \mcc{0.35 (fixed)}               & \mcc{0.35 (fixed)}               & \mcc{0.21 (fixed)}                       \\
			Sum of the fractional radii          & $r_{\rm A}+r_{\rm b}$     & \mcc{$0.1076 \pm 0.0081$}        & \mcc{$0.1019 \pm 0.0080$}        & \mcc{$0.0957 \pm 0.0043$}        \\
			Ratio of the radii                   & $k = R_{\rm b}/R_{\rm A}$ & \mcc{$0.1006 \pm 0.0047$}        & \mcc{$0.1148 \pm 0.0057$}        & \mcc{$0.1922 \pm 0.0078$}        & 0.1079  & 0.0011      \\
			Inclination (\degrees)               & $i$                       & \mcc{$86.77 \pm 0.78$}           & \mcc{$87.46 \pm 0.88$}           & \mcc{$88.96 \pm 0.88$}           & 85.56   & 0.22        \\
			Fractional radius of the star        & $r_{\rm A} = R_{\rm A}/a$ & \mcc{$0.0977 \pm 0.0070$}        & \mcc{$0.0914 \pm 0.0068$}        & \mcc{$0.0803 \pm 0.0033$}        & 0.1078  & 0.0027      \\
			Fractional radius of the star        & $r_{\rm b} = R_{\rm b}/a$ & \mcc{$0.0098 \pm 0.0011$}        & \mcc{$0.0091 \pm 0.0013$}        & \mcc{$0.0154 \pm 0.0011$}        \\
			\hline
			Stellar mass (\Msun)                 & $M_{\rm A}$               & 1.090    & 0.048    & 0.017      & 1.089    & 0.047    & 0.017      & 0.792    & 0.058    & 0.280      & 1.200   & 0.041       \\
			Stellar radius (\Rsun)               & $R_{\rm A}$               & 1.221    & 0.090    & 0.006      & 1.142    & 0.085    & 0.006      & 0.903    & 0.046    & 0.011      & 1.392   & 0.044       \\
			Stellar surface gravity (cgs)        & $\log g_{\rm A}$          & 4.302    & 0.063    & 0.002      & 4.360    & 0.066    & 0.002      & 4.426    & 0.037    & 0.005      & 4.231   & 0.020       \\
			Stellar density (\psun)              & $\rho_{\rm A}$            & \mcc{$0.60 \pm 0.13$}            & \mcc{$0.73 \pm 0.17$}            & \mcc{$1.08 \pm 0.13$}            & 0.447   & 0.033       \\
            Age (Gyr)                            & $\tau$                    & \ermcc{5.0}{0.8}{1.6}{0.6}{0.8}  & \ermcc{3.6}{1.9}{2.9}{0.8}{2.2}  & \ermcc{16.1}{1.1}{5.2}{3.9}{5.7} & 
            \multicolumn{2}{c}{$7^{+2}_{-1}$} \\
            \hline
            Planet mass (\Mjup)                  & $M_{\rm b}$               & 0.291    & 0.017    & 0.003      & 0.378    & 0.022    & 0.004      & 1.30     & 0.19     & 0.03       & 0.311   & 0.017       \\
			Planet radius (\Rjup)                & $R_{\rm b}$               & 1.20     & 0.14     & 0.01       & 1.28     & 0.15     & 0.01       & 1.69     & 0.12     & 0.02       & 1.462   & 0.059       \\
			Planet surface gravity (\mss)        & $g_{\rm b}$               & \mcc{$5.0 \pm 1.2$}              & \mcc{$5.8 \pm 1.5$}              & \mcc{$11.3 \pm  2.2$}            & 2.530   & 0.036       \\
			Planet density (\pjup)               & $\rho_{\rm b}$            & 0.159    & 0.059    & 0.001      & 0.170    & 0.065    & 0.001      & 0.252    & 0.064    & 0.003      & 0.099   & 0.012       \\
			Equilibrium temperature (K)          & \Teq\                     & \mcc{$1326 \pm   52$}            & \mcc{$1282 \pm   52$}            & \mcc{$1013 \pm   54$}            & 1379    & 31          \\
			Orbital semimajor axis (AU)          & $a$                       & 0.0581   & 0.0009   & 0.0003     & 0.0581   & 0.0008   & 0.0003     & 0.0523   & 0.0013   & 0.0006     & 0.0600 & 0.0007       \\
			\hline
		\end{tabular}
	\end{table}
\end{longrotatetable}

\begin{table}
	\centering
	\caption{Predicted transit depths in various filters for the `planet transits A' and `planet transits B' scenarios, based on the $r$-band transit depth and the dilution as a function of wavelength. SDSS $ugriz$ and Bessel $UBVRI$ filter curves from the Nordic Optical Telescope,\tablenotemark{a} $JHK_{\rm s}$ filter curves from SPHERE.\tablenotemark{b}} \label{tab:fluxRatios}
	\begin{tabular}{l c c }
		\hline \hline
		Filter & Depth transiting A (\%) & Depth transiting B (\%) \\
		\hline
		$u$         & 1.324 & 0.569 \\
		$g$         & 1.227 & 0.930 \\
		$r$         & 1.164 & 1.164 \\
		$i$         & 1.132 & 1.285 \\
		$z$         & 1.105 & 1.385 \\
		$U$         & 1.323 & 0.574 \\
		$B$         & 1.263 & 0.795 \\
		$V$         & 1.199 & 1.035 \\
		$R$         & 1.158 & 1.189 \\
		$I$         & 1.124 & 1.316 \\
		$J$         & 1.066 & 1.532 \\
		$H$         & 1.016 & 1.718 \\
		$K_{\rm s}$ & 1.005 & 1.757 \\
		\hline
	\end{tabular}
	\tablenotetext{a}{Filter IDs 1-5 and 82-86, \url{http://www.not.iac.es/instruments/filters/filters.php}}
	\tablenotetext{b}{\url{https://www.eso.org/sci/facilities/paranal/instruments/sphere/inst/filters.html}}
\end{table}

\end{document}